\documentstyle[preprint,aps,floats,tighten]{revtex}
\begin{document}
\draft\preprint{titcmt-95-1,adap-org/9501003}
\title{A set of hard spheres with tangential inelastic collision
as a model of granular matter:
$1/f^\alpha$ fluctuation, non-Gaussian distribution, and
convective motion}
\author{Y-h. Taguchi\cite{aaa}}
\address{Department of Physics, Tokyo Institute of Technology,
Oh-okayama,  Meguro-ku, Tokyo 152, Japan}
\author{Hideki Takayasu}
\address{Graduate School of Information Sciences, Tohoku University,
Aoba-ku, Sendai 980, Japan}
\date{\today}
\maketitle
\begin{abstract}
A set of hard spheres with tangential inelastic collision
is found to reproduce observations of real and numerical
granular matter.
After time is scaled so as to cancel energy dissipation due to
inelastic collisions out,
inelastically colliding hard spheres in  two
dimensional space
come to have
$1/f^\alpha$ fluctuation of total energy,
non-Gaussian distribution of displacement vectors, and
convective motion of spheres, which
hard spheres with elastic collision,
a conventional model of granular matter, cannot reproduce.
\end{abstract}
\narrowtext
Granular matter attracted
attention of physicists  recently\cite{review}.
It exhibits many strange behaviors:
segregation\cite{seg}, surface standing wave\cite{wave},
bubbling\cite{buble}, $1/f^\alpha$ fluctuation\cite{1/f},
convection\cite{conv}, non-Gaussian distribution\cite{ichiki,non-G},
and turbulence\cite{ichiki,turb}.
Although  numerical simulations
easily reproduce most of them,
we cannot understand their mechanisms.
This is because we do not have
any universal model which can exhibit
these strange behaviors observed in dynamics of granular matter.
The lack of universal model prevents us from
going beyond researching
individual phenomena.

One of granular models used most frequently,
other than conventional continuous body
approximations\cite{buble,cont},
is a set of hard spheres colliding
with each other (kinetic theory\cite{Campbel}).
In kinetic theory,
a set of hard spheres
is assumed to be in thermal equilibrium,
thus probability distribution function (PDF)
of hard spheres' velocity
should be Gaussian with
variance  proportional to temperature.
Due to this analogy,  PDF which  granular particle velocity obeys
is also assumed to be Gaussian,
although none observed it directly,
and kinetic energy of granular particle
is called 'granular temperature'.
However, we do not think that
this analogy is suitable one for dense granular state,
which corresponds to weakly fluidized granular matter;
e.g., vibrating bed of powder,
gas fluidized bed, chute flow, and hopper flow.

First, probability distribution function  of
particle velocity deviates from
Gaussian\cite{ichiki,non-G}.
Second,
kinetic theory cannot explain
the frequent appearance of $1/f^\alpha$ fluctuations\cite{1/f}.
Third, it  cannot also explain why fluidized granular matter
behaves like fluid.
Cooperative macroscopic fluid mode
should appear only in the limit that
the number of particles is large when we employ kinetic theory.
However, fluid like motion
can appear in granular matter composed of
 less than a thousand particles\cite{conv}.

The purpose of this letter is
to modify hard sphere model so as to
improve insufficiencies mentioned above.
We show that considering
 tangential inelastic collision and
scaling velocity
give rise to the appearance of $1/f^\alpha$ fluctuations,
which also explains why non-Gaussian PDF can appear.
Furthermore,  tangential friction allows
a set of hard spheres to have
cooperative modes of fluid motion,
even if the number of spheres is
a few hundreds.

Let us consider
why kinetic theory cannot describe
dense granular state.
It is a natural assumption that
each particle is a hard sphere.
Main difference between
a set of hard spheres and granular matter is
whether the energy is conserved or not.
The application of kinetic theory to granular matter
is supported by the assumption
that granular matter can control amount of dissipation so as to cancel out
energy input and can converge to steady state.
In order to know whether this assumption really stands,
 we investigate
 a set of
hard spheres with inelastic collision in the following.

First, let us reconsider how to get steady state.
Although one may believe hard spheres
with inelastic collision
cannot have steady state without energy input,
it is not true from the theoretical point of view.
If we rescale velocity properly,
there can exist steady state.
For example, let us consider a hard sphere moves
along segment of length $L$.
The collisions between the sphere and ends are inelastic
(the coefficient of restitution $e  < 1$).
Suppose that the sphere is launched
 from one end $A$ towards the other end $B$ at time $t=0$
with the velocity $v$.
Time $T$ when the sphere reaches  the  end $B$ is $L/v$.
After the sphere collides with the  end $B$,
 the velocity $v$ decreases to  $ev$,
and
the sphere reflected by the end  $B$ will come back,
at time $L/v+L/{ev}$,
to the end $A$.
Hence  until
the sphere loses its velocity completely it takes
$L/v + L/{ev} + L/{e^2v} + \cdots$ which
diverges, since $e<1$.
Therefore energy of
the sphere never vanish within finite period.
What we observe here is only slowing down of motion.
Rescaling time by the factor of $e$
every time the sphere collides with one of the ends,
we can have steady state:
a sphere moves with constant velocity $v$.
In the following, we regard this scaled state
as a model of granular matter instead of
employing kinetic theory.
In this scaled state
number of collisions $n_{col}$
is employed as a unit of time.

The scaled state can be defined even in
higher spatial dimensions.
First, we pack
hard spheres with inelastic collision into a $d$-dimensional box
and calculate mean energy dissipation rate
$\varepsilon = - \langle d E /dn_{col} \rangle$,
where $E$ is total kinetic energy
and average is taken over long period.
Dealing with $v \exp( \sqrt{\varepsilon} n_{col})$ instead of
$v$, we can have scaled state
in higher spatial dimensions.
In the following, we describe physical
variables in this scaled state
by  the character with tilde.
For example, energy in scaled state is
$\tilde{E}= E \exp( \varepsilon n_{col})$, and
velocity is $\tilde{v}=v \exp( \sqrt{\varepsilon} n_{col})$.

In the following scaled state in 2D,
in which we use circles instead of sphere
to represent each particle,
is considered.
Here we consider only inelasticity
of collision along tangential direction.
(Here normal vector  when they collide is
defined as component along line passing through
centers of two colliding circles,
and tangential vector is perpendicular to it.)
Normal components of velocity of each particle
simply changes its direction oppositely due to collision.
This is because we would like to make
the model as simple as possible.
When this definition is too simple to
reproduce behaviors which we would like to consider,
additional mechanism, e.g., inelastically of collision
along normal direction, may have to be taken into account.

\begin{figure}
\begin{center}
\input{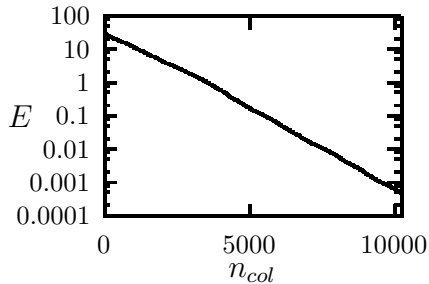}
\caption{Time development of energy $E$}
\label{fig:eng_diss}
\end{center}
\end{figure}

\begin{figure}
\begin{center}
\input{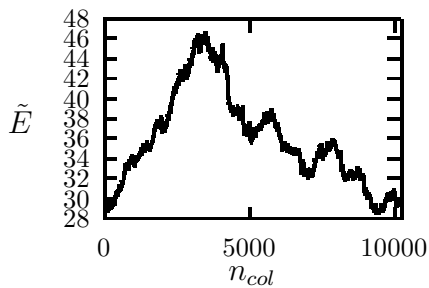}
\caption{Scaled energy $\tilde{E}$ in ideal granular matter}
\label{fig:eng_fluct}
\end{center}
\end{figure}

\begin{figure}
\begin{center}
\input{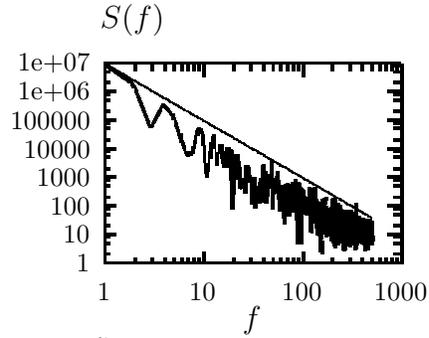}
\caption{Fourier power spectrum of $\tilde{E}(n_{col})$.
Data points are sampled each ten collisions from
time sequential data,
(thus data length $= 256 \times 40 /10 =1024$) and
power spectrums are averaged over obtained ten samples.
Straight line indicates $1/f^2$ dependence upon
frequency $f$}
\label{fig:1/f}
\end{center}
\end{figure}

For simulating 2D system, we put hard spheres
into a triangular box with elastic wall.
By changing area of rectangular,
we control averaged density of hard spheres.
The rectangular has the width of $(2l_x+1) a \phi$ and the
hight of $2\sqrt{3}(l_y-1) a \phi$,
where $a (=1.0)$ is radius of each circle and
$\phi$ is expansion rate.
When $\phi=1$, granular matter forms tight rectangular lattice
(See Fig.\ref{fig:rect}).
\begin{figure}
\vspace{3.5cm}
\caption{Schematic of numerical setup.
Bold rectangular represents dense state with $l_x=l_y=6$.
Broken rectangular indicates expanded state with $\phi \simeq 1.3$.}
\label{fig:rect}
\end{figure}
Thus it
quantizes how much the rectangular expands comparing with
dense packed state.
Hereafter, $y$ axis is taken to be parallel to vertical direction and
$x$ is horizontal.
Starting from initial condition
that center of each circle forms uniformly elongated triangular lattice
and velocities $v_x$ and $v_y$ are taken from uniform
random number $\in [-0.5,0.5]$,
we simulate the system
composed of $256$ circles
($l_x=l_y=16$, tangential coefficient of restitution $e=0.5$, and
$\phi=1.1$).
Figure \ref{fig:eng_diss} shows the
time development of total energy $E =\sum_i (v_{ix}^2+v_{iy}^2)/2$,
where $i$ denote the numbering of circles.
$E$ monotonically decreases exponentially as a function
of $n_{col}$.
Using the data, we compute energy dissipation rate
$\varepsilon = -\frac{1}{n_{col}^{tot}} \sum_{n_{col}}
 \log [E(n_{col}+1)/E(n_{col})]$,
where $n_{col}^{tot}$ is total observation time ($=40 \times 256$), and
plot scaled energy $\tilde{E}=E \exp(\varepsilon n_{col})$
in Fig.\ref{fig:eng_fluct}.
$\tilde{E}$ fluctuate violently, and
the amount of fluctuation is not ignorable comparing
with mean value $\langle \tilde{E} \rangle$.
The Fourier power spectrum
of $\tilde{E}(n_{col})$ is shown in Fig.\ref{fig:1/f}.
It has power law dependence upon
frequency $f$, and the exponent is close to $-2$.
Thus, $\tilde{E}$ fluctuates as random walker does
(i.e., the fluctuation is not stationary.).
This gives rise to the conclusion that
scaled state is not stationary.
Thus introduction of dissipation destroys basic
assumption of kinetic theory,
stationarity of granular matter.

Now let us consider real granular matter.
In real granular matter,
energy dissipation is believed to be balanced with
energy inputs, e.g., gravity, shear force, or
drug force from fluid.
However, these energy inputs are usually stationary
so that they cannot suppress the
instationary fluctuation completely.
This means, even in real granular matter,
we will  observe the instationary fluctuation
observed in scaled state.
Perhaps this is the reason why
we frequently observe $1/f^\alpha$ fluctuation in granular matter
over the wide range of phenomena.

\begin{figure}
\begin{center}
\input{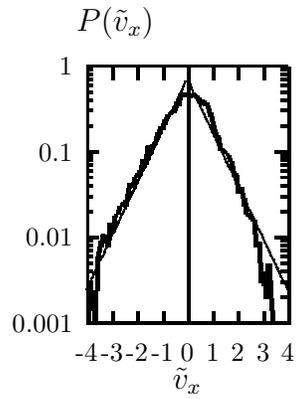}
\caption{PDF of scaled velocity $\tilde{v}_x$.
Thin line is exponential distribution.}
\label{fig:PDF}
\end{center}
\end{figure}

Next we consider deviation of velocity PDF from
Gaussian.
Figure \ref{fig:PDF} shows PDF of scaled velocity $\tilde{v}_x$.
It is calculated from the system having $256$ circles
 and averaged over $40$ snap shots
taken every $256$(=number of circles) collisions.
It  deviates from Gaussian,
 and its form reminds us non-Gaussian PDF observed
in numerical simulation of vibrating bed of powder.
In contrast to this,
a set of elastic spheres (i.e.,$e=1$) simulated by our code
gives us Gaussian PDF of velocity.

Here we can explain
why scaled state can have non-Gaussian PDF.
As shown above, the fluctuation
of scaled energy $\tilde{E}$ behaves as fluctuation of
the position of random walkers whose motion is unbounded.
This means, energy fluctuation is unbounded,
thus, the fluctuation of velocity is not bounded, either.
It prevents scaled state from having
PDF of velocity with
finite variance which
Gaussian PDF must have.
Therefore PDF of velocity in scaled state
cannot obey Gaussian.
It is very suggestive that
exponent of power PDF observed in numerical simulation
of vibrated bed of powder is very close to
marginal value $-3$,
the boundary between PDF with finite variance and
that with infinite variance.
(When PDF of $v$ obeys power law PDF as $P(v) \sim v^{-\beta}$,
finite variance can exist only when $\beta>3$.)
This explanation also coincides with the results obtained by
Hayakawa and Ichiki\cite{ichiki}, who
found deviation of particle velocity PDF from Gaussian
accompanied with $1/f^2$ fluctuation in numerical fluidized bed.

Finally we check whether scaled state can
exhibit the cooperative dynamics.
When we consider instantaneous scaled velocity,
we could not find any spatial
structures.
However, when we consider
displacement vectors over
long period
$\Delta {\bf x}_i^{\delta n}(n) = {\bf x}_i(n+\delta n)-{\bf x}_i(n)$,
we can observe cooperative spatial structure.
Figure \ref{fig:conv} shows $\Delta {\bf x}_i^{\delta n}$
with $\delta n=20 \times 256$.
\begin{figure}
\vspace{4cm}
\caption{Spatial structure of displacement vectors
$\Delta {\bf x}_i^{\delta n}(n) $}
\label{fig:conv}
\end{figure}
We can see eddy like structure, which may
give rise to convection observed in real vibrated bed of powder.
Thus, cooperative motion can be
observed only when
motion is coarse grained during long period.
The reason why we observed
eddy
even in instantaneous velocity
using soft-core potential model\cite{turb}
can be understood in this consequence.
In the soft-core potential model,
each collision should  be regarded as
not individual collisions but
coarse grained collisions.
This interpretation of soft-core potential model
is supported by recent findings
where convection
 disappears in the limit of zero collision time
in soft-core potential model\cite{crit_duran}.
The zero collision time limit in soft-core potential model
clearly does not converge to the realistic situation.
One should not take zero collision time limit since
each collision in soft-core potential model
should be regarded as a set of several collisions
in real granular matters.

Finally let us discuss hierarchical structure
of scaled state.
Consider a granular particle
surrounded by six neighboring particles
fixed on the lattice points of triangular lattice (Fig.\ref{fig:biliard}).
\begin{figure}
\vspace{2.5cm}
\caption{A granular particle as a dispersing billiard.
Surrounding six particles are fixed.
Broken curve show the region where
center of shaded particle  can move around.}
\label{fig:biliard}
\end{figure}
The motion of central movable particle
is nothing but that of dispersing billiard\cite{sinai}.
The dispersing billiard is known to be
chaotic and Ergodic.
Thus, we can expect  short-time motion of individual particles
is chaotic, although this analogy is not exact since
in granular material surrounding particles can move and
collide with each other inelastically.
However, we found that
Fourier power spectrum
obtained from time sequential data of scaled velocity $\tilde{v}_i$
obeys Lorentzian.
This means, auto correlation of scaled velocity of individual
particle decays exponentially as time proceeds.
This decay indicates that dynamics in scaled state
still remains chaotic.
Furthermore,
we cannot distinguish Fourier power spectrum observed in
system with elastic collision from that with inelastic collision.
Thus, introducing inelastic collision does not seem to suppress
chaotic motion of individual particle.

In contrast to chaotic motion of each particle,
averaged quantity has long ranged or long time
correlations.
Scaled energy $\tilde{E}$, which is a sort of
spatial averaged velocity, has
long time correlation, and
displacement vector $\Delta {\bf x}_i^{\delta n}$,
which is temporal averaged velocity,
has long range correlation.
Thus, we can conclude that
macroscopic and/or long time behaviors
originates in averaging procedure.
What Goldhirsch and Zanetti found\cite{goldhirsch}
may support this conclusion since
they have found long range correlation
in inelastically colliding hard spheres
after the local velocity was coarse-grained.

Finally, one should note that
we do not intend to insist that
scaled state can exhibit all
behaviors which real granular material shows.
Our modeling corresponds to
ideal gas which explains equation of state correctly,
but reproduction of  phase transition
needs van der Waals correction to it.
Our modeling  may also need to be modified
to explain other behaviors which are not explained in this letter.

In summary, we have proposed a set of hard spheres with
tangential inelastic collision as
a model of dynamics of granular material.
Scaled state of a set of hard spheres exhibits
$1/f^\alpha$ fluctuation, non-Gaussian PDF, and
eddy like flow, all of which
are observed numerically and/or experimentally in real granular materials.
Using analogy to the scaled state,
we suggest that there will be no stationary state in
dynamical state of granular materials,
and macroscopic motion comes from
averaging the chaotic motion of
individual particles.
HT thanks Dr. M. Takayasu for helpful discussions.
YHT thanks people who developed Linux on which all calculations
were performed.

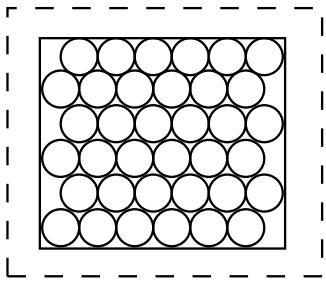
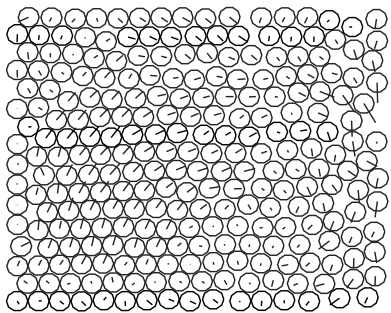
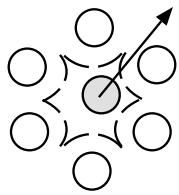
\end{document}